\documentclass[12pt]{iopart}
\usepackage{iopams} 

\usepackage{braket}

\usepackage{color}
\usepackage{epsfig}
\usepackage{epstopdf}
\usepackage{float}
\usepackage[ruled,linesnumbered]{algorithm2e}
\usepackage[colorlinks,linkcolor=magenta,anchorcolor=blue,citecolor=blue,urlcolor=blue]{hyperref}
\usepackage{graphicx, caption}

\def\L{{\cal{L}}}
\def\D{{\cal{D}}}
\def\d{{\Delta}}

\newcommand*{\dyad}[1]{ |{#1}\rangle\langle{#1}| }
\newcommand*{\expval}[1]{\langle{#1}\rangle}
\newcommand*{\pdv}[2]{\frac{\partial #1}{\partial #2}}

\usepackage{cite}

\begin{document}

\title{Quantum hypothesis testing via robust quantum control}

\author{Han Xu$^{1,2}$, Benran Wang$^3$, Haidong Yuan$^{4,*}$ and Xin Wang$^{1,*}$}
\address{$^1$ Department of Physics, City University of Hong Kong, Tat Chee Avenue, Kowloon, Hong Kong SAR, China, and City University of Hong Kong Shenzhen Research Institute, Shenzhen, Guangdong 518057, China}
\address{$^2$ School of Physics and Technology, Wuhan University, Wuhan, Hubei 430072, China}
\address{$^3$ School of Computer Science and Engineering, Sun Yat-Sen University, Guangzhou, Guangdong 510006, China}
\address{$^4$ Department of Mechanical and Automation Engineering, The Chinese University of Hong Kong, Shatin, Hong Kong SAR, China}
\ead{hdyuan@mae.cuhk.edu.hk and x.wang@cityu.edu.hk}
\vspace{10pt}
\begin{indented}
\item[]September 2023
\end{indented}

\begin{abstract}
Quantum hypothesis testing plays a pivotal role in quantum technologies, 
    making decisions or drawing conclusions about quantum systems based on observed data.
    Recently, quantum control techniques have been successfully applied to quantum hypothesis testing, enabling the reduction of error probabilities in the task of distinguishing magnetic fields in  presence of environmental noise. In real-world physical systems, such control is prone to various channels of inaccuracies. Therefore improving the robustness of quantum control in the context of quantum hypothesis testing is crucial.
    In this work, we utilize optimal control methods to compare scenarios with and without accounting for the effects of signal frequency inaccuracies.
    For parallel dephasing and spontaneous emission, the optimal control inherently demonstrates a certain level of robustness, while in the case of transverse dephasing with an imperfect signal, it may result in a higher error probability compared to the uncontrolled scheme.
    To overcome these limitations, we introduce a robust control approach optimized for a range of signal noise, demonstrating superior robustness beyond the predefined tolerance window.
    On average, both the optimal control and robust control show improvements over the uncontrolled schemes for various dephasing or decay rates, with the robust control yielding the lowest error probability.
\end{abstract}

%
\vspace{2pc}

\section{Introduction}

Quantum hypothesis testing is a statistical methodology that combines the principles of quantum mechanics and hypothesis testing. It involves making decisions or drawing conclusions about quantum systems based on observed data. 
Quantum hypothesis testing, in contrast to classical hypothesis testing, considers superposition and entanglement, requiring identification of quantum states through measurement and subsequent classical processing of the measurement outcomes.
By comparing different hypotheses for the measurement outcomes, it enables determination of the most probable hypothesis. 
Quantum hypothesis testing has been regarded as one of the foundational tasks in various fields of quantum information science \cite{hayashi2016quantum,watrous2018theory} including quantum communication \cite{wang2012one,Matthews2014Finite,cheng2022simple}, quantum illumination \cite{Lloyd2008Science,Tan2008Quantum,Wilde2017Gaussian,yung2020one}, and quantum channel discrimination \cite{aharonov2002measuring,tsang2012continuous,chen2019zeroerror,bergh2023composite}. 

Quantum control has emerged as a crucial tool in recent years for advancing quantum technologies \cite{Glaser2015Cat,Koch2016Control,boscain2021intro}. Its applications extend to fields such as quantum computation \cite{Palao2002Quantum,Calarco2004Quantum} and quantum simulation \cite{Doria2011Optimal,Cui2017Optimal,Omran2019Generation}, empowering breakthroughs in these fields. In contrast to the quantum computation problem, where the objective is to drive the initial state towards a target state as closely as possible \cite{palao2003optimal,khaneja2005optimal}, hypothesis testing with quantum control focuses on enhancing the distinguishability between quantum states \cite{helstrom1976}. This parallels the concept of quantum parameter estimation \cite{liu2017quantum,liu2017control,liu2019quantum}. Within this context, the figure of merit, when treated as a functional of the control, serves as a measure of the distinguishability between the quantum states. By numerically extremizing this functional, one maximizes the distinguishability between states as outcomes from different hypotheses, thus minimizing the error in quantum hypothesis testing.

However, in practice, the accuracy of the hypothesis in a quantum system can be compromised by factors such as uncertainties associated with the parameters in the system Hamiltonian. Ensuring control robustness in the face of such inaccuracies becomes crucial \cite{Glaser2015Cat}. 
While quantum control has demonstrated success in discriminating states \cite{higgins2009mixed} and dynamics \cite{basilewitsch2020optimally} of systems, only a few studies have addressed the issue of robustness. 
Recent works have examined the impact of uncharacterized state preparation noise on state discrimination \cite{flatt2019multiple,Dalla2020Quantum}, but the effects of imperfect signal noise in quantum hypothesis testing have not been adequately explored. 
Robust quantum control, capable of mitigating the impact of signal noise, holds the potential to improve distinguishability in the face of uncertainties in Hamiltonian parameters. This is a common situation encountered in quantum parameter estimation challenges \cite{young2009optimal,chakrabarti2012optimal}.
In this context, it's important to note that both state preparation noise and signal noise are not environmental noise. Environmental noise results from interactions with the surrounding environment and affects error probabilities, specifically in the case of dephasing dynamics as explored in later part of this paper. On the other hand, state preparation noise and signal noise pertain directly to the system under investigation.
In this paper, we define robustness in terms of the average distinguishability within a predefined tolerance window for signal noise. We employ quantum control techniques to enhance the average distinguishability under different types of environmental noise and compare the robustness of various pulse profiles.

The remainder of this paper is structured as follows. \Sref{sec:preliminaries} provides an overview of binary hypothesis testing with quantum control in the presence of dephasing dynamics and spontaneous emission. In \Sref{sec:method}, we outline the optimization methods employed to address the optimal control problem in this context. Moving on to \Sref{sec:results_}, we apply the optimal control techniques to the discrimination of magnetic fields under various types of environmental noise. The key outcomes of this study, focusing on the robustness of hypothesis testing with quantum control, are also presented. Finally, we conclude our findings in \Sref{sec:conclusion}.


\section{Preliminaries}\label{sec:preliminaries}

\begin{figure}[t]
    \centering
    \captionsetup{width=\linewidth}
    \includegraphics[width=0.98\linewidth]{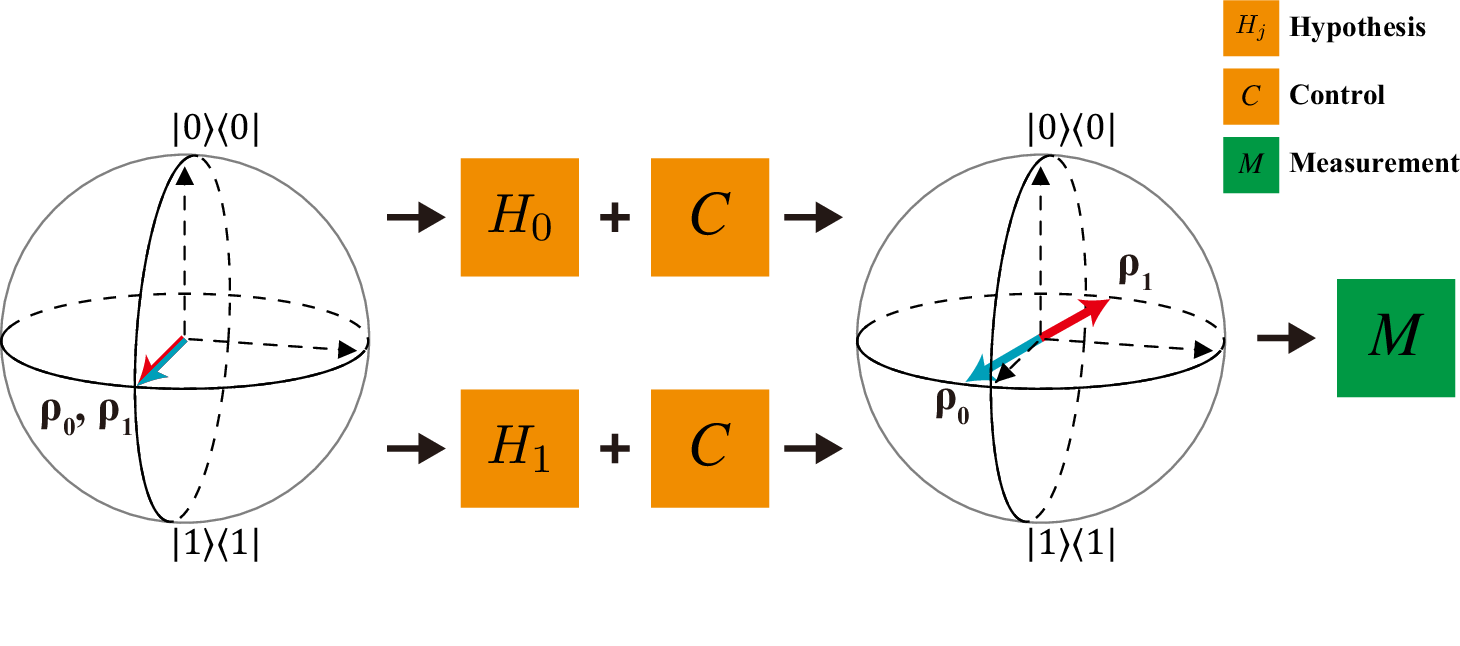}
    \caption{Schematics of the binary hypothesis testing with quantum control for the detection of an unknown magnetic field in a qubit system. The null hypothesis ($H_0$) represents the background, while the alternative hypothesis ($H_1$) signifies the presence of the magnetic field. A control ($C$) is applied to enhance the distinguishability of states $\rho_0$ and $\rho_1$, allowing them to evolve under the respective influences of $H_0$ and $H_1$.
    }\label{fig:hypo}
\end{figure}


In hypothesis testing, we assign the proposition of the system being in state $j$ to the hypothesis $H_j$, where $j$ denotes the index of the hypothesis. The binary hypothesis testing problem involves deciding between two hypotheses: the null hypothesis $H_0$ representing the background and the alternative hypothesis $H_1$ representing the signal, namely $j\in\{0,1\}$. To illustrate, consider the detection of an unknown magnetic field in a qubit system as shown in \fref{fig:hypo}. The two hypotheses in this binary hypothesis testing problem correspond to two Hamiltonians $H_0=0$ and $H_1=\vec{B}\cdot\vec{\sigma}$ with $\vec{B}$ denoting the magnetic field and $\vec{\sigma}$ defined by the Pauli matrices $\vec{\sigma}=(\sigma_x, \sigma_y, \sigma_z)$.

In the process of identifying the correct hypothesis, two types of errors can occur. The \emph{type I} (or \emph{false positive}) error corresponds to the probability of accepting a false alternative hypothesis $H_1$ when the null hypothesis $H_0$ is true. Conversely, the \emph{type II} (or \emph{false negative}) error represents the probability of accepting a false null hypothesis $H_0$ when the alternative hypothesis $H_1$ is true. Although the consequences of wrongly accepting or rejecting either hypothesis can differ significantly, often associated with asymmetric \textit{a priori} probabilities assigned to each hypothesis, this paper focuses on the symmetric scenario.

In the case of Markovian environments, $H_0$ and $H_1$ under consideration govern the time evolutions of the system. These dynamics can be described by the master equation of Lindblad type \cite{breuer2002},
\begin{equation}
    \partial_t\rho(t) = \L[\rho(t)].
    \label{eq:me0}
\end{equation}
For unitary evolution, the dynamics are described by the equation $\L[\rho] = -i [H,\rho]$. However, for noisy evolution, additional terms arise due to the interaction with the environment. The Lindblad master equation captures this noisy process, given by $\L[\rho] = -i [H,\rho] + \Gamma[\rho]$, where $\Gamma$ represents the contribution from the interaction with the environment.
In this paper, we shall consider the environmental noise of dephasing and spontaneous emission.

In the context of a quantum control system, the Hamiltonian is defined as $H = H_{j} + \sum_{k=1}^{K}u_{k}(t)H_{c,k}$, where $u_k(t)$ represents the control amplitude, $H_{c,k}$ describes the control Hamiltonian, and $K$ is the total number of control channels.
This formulation allows us to modulate the system's dynamics by applying the same control field to both $H_0$ and $H_1$.

In our approach, we consider a quantum control system that employs piecewise constant pulse sequences within the time interval $[0,T]$. To ensure the validity of the Markovian approximation under the quantum control, we assume that the evolution described by the master equation in \eref{eq:me0} is much slower than the decay of correlations in the environment \cite{breuer2002}.
Consequently, the time evolution of the density matrix can be expressed as
\begin{equation}\label{eq:rho_evolution}
    \rho(T) = \prod_{n=0}^{N-1}e^{\d t \L_n}\rho(0),
\end{equation}
where $\rho(0)$ denotes the initial state, $N=T/\d t$ represents the total number of time slices, and $\L_n$ is the superoperator associated with the $n$th time slice. 

Let $\rho_0(T)$ and $\rho_1(T)$ denote the time-evolved states under $H_0$ and $H_1$, respectively. The goal of optimizing the control field is to maximize the distinguishability between $\rho_0(T)$ and $\rho_1(T)$, making them as discernible as possible.

When the initial state is predetermined, the problem of discriminating $H_0$ and $H_1$ reduces to the task of discriminating quantum states through measurement at the conclusion of the evolution. 
The measurement which minimizes the probability of incorrectly identifying the state is known as the Helstrom measurement \cite{yuen1975,helstrom1976,holevo2011}.
However, implementing the Helstrom measurement can be challenging as it typically is defined with the time-evolved states and thus depends on the dynamics of the system, adding complexity to its physical realization.
Moreover, when multiple copies of the state are available, finding a practical implementation of the Helstrom measurement becomes even more difficult. 
Consequently, researchers have dedicated efforts to develop local measurement strategies that can achieve comparable performance to Helstrom measurements \cite{acin2005multiple,higgins2011multiple,vargas2021quantum}. In the following sections, we define the error probabilities associated with these two types of measurements in the context of hypothesis testing.

\textit{Fixed local measurement.} 
Assuming that each quantum system is measured using a fixed apparatus, the \emph{type I} and \emph{type II} errors can be expressed as $p(1|H_0)=\tr(\rho_0E_1)$ and $p(0|H_1)=\tr(\rho_1E_0)$, respectively. Here, the positive operator-valued measure (POVM) element $E_j$ is associated with the  outcome indicating that hypothesis $H_j$ is true. The error probability $P_e$ is defined as
\begin{equation}\label{eq:obj_func_fix}
    P_e = \pi_0\tr(\rho_0E_1) + \pi_1\tr(\rho_1E_0),
\end{equation}
where $\pi_0$ and $\pi_1$ represent the prior probabilities of each hypothesis. In the case of symmetric scenario, we have $\pi_0=\pi_1=1/2$.

\textit{Helstrom measurement.}
Upon minimizing the error probability across all possible POVM $\{E_0,E_1\}$, the Helstrom bound provides the lowest achievable error probability \cite{audenaert2007discriminating,calsamiglia2008quantum}, given by
\begin{equation}
    P_e^{H}=\frac{1}{2}\left(1-\D_{\tr}\left(\rho_{0}, \rho_{1}\right)\right),
\end{equation}
where the trace distance $\D_{\tr}$ is a quantum generalization of the classical trace distance and serves as a metric for quantifying the distance between two quantum states. It is defined as
\begin{equation}
   \D_{\tr}\left(\rho_{0}, \rho_{1}\right) \equiv \frac{1}{2}\left\|\rho_{0}-\rho_{1}\right\|_{1},
\end{equation}
with $\|X\|_1 \equiv \tr\,[\sqrt{X^\dagger X}]$ representing the \emph{trace norm} \cite{nielsen2010}. A smaller value of $\D_{\tr}$ indicates a closer proximity between the two quantum states. Specifically, when $\D_{\tr}(\rho_1, \rho_2)=1$ the states are orthogonal, enabling perfect discrimination. On the other hand, if $\D_{\tr}(\rho_1, \rho_2)$ is equal to 0, it implies that the states cannot be discriminated. Achieving the Helstrom bound involves a projective measurement onto the eigenstates of the matrix $\rho_0-\rho_1$ \cite{holevo2011}. It is important to note that since the density matrices evolve under $H_0$ and $H_1$, the Helstrom measurement is inherently time-dependent.


In general, there exist various approaches to reduce the error probability when the final states are nonorthogonal. One strategy involves preparing $n$ identical copies of the initial states, allowing them to evolve under the respective Hamiltonians, and performing measurements at the desired target time. The error probability can decrease exponentially with $n$ \cite{chernoff1952measure}, and this can be facilitated through techniques such as collective measurement \cite{audenaert2007discriminating,calsamiglia2008quantum}, adaptive local measurement \cite{acin2005multiple,higgins2011multiple} and sequential strategy with fixed local measurement \cite{vargas2021quantum}.
Alternatively, the error probability can be mitigated by dynamically controlling the evolution through quantum control operations \cite{basilewitsch2020optimally}. This approach enables the manipulation of the system's dynamics to optimize the distinguishability between states. Remarkably, even a zero error probability becomes feasible by employing suitable control operations, both in the context of unitary evolution \cite{aharonov2002measuring} and in the presence of noisy dynamics \cite{chen2019zeroerror}.


\section{Method}\label{sec:method}
In this section, we provide a concise overview of two methods utilized in this work: the Gradient Ascent Pulse Engineering (GRAPE) method \cite{khaneja2005optimal,basilewitsch2020optimally} and the simulated annealing GRAPE (SAGRAPE) method \cite{ram2022robust,mahesh2022quantum}.

In general, the GRAPE method is employed to automatically determine the optimal piecewise constant control that minimizes the figure of merit, $P_e$ or $P_e^{H}$, according to their gradients with respect to the control amplitude $u_k$.
On the other hand, the SAGRAPE method combines the gradient ascent technique with simulated annealing, a stochastic global search optimization algorithm. This combined approach enables faster convergence and helps overcome local optima that may arise in optimal control problems \cite{mahesh2022quantum}. 
The pseudocode for the aforementioned algorithms is presented below.

Let's consider the numerical optimization of the Helstrom bound as an illustrative example. To simplify the computation, we introduce an objective function,
\begin{equation}
    \D(\rho_0, \rho_1) = \frac{1}{2}\tr(|\rho_0 - \rho_1|^2).
    \label{eq:obj_func}
\end{equation}
This objective function allows us to analyze the gradients of polynomial functions of $\rho_0$ and $\rho_1$ \cite{xu2004optimal,basilewitsch2019quantum}. 

The objective function \eref{eq:obj_func} provides a lower bound for the trace distance $\D_{\tr}(\rho_0, \rho_1)$, i.e.~$\D(\rho_0, \rho_1)\le\D_{\tr}(\rho_0,\rho_1)$, indicating that maximizing $\D$ guarantees an increase in the desired quantity $\D_{\tr}$ and a decrease in the error $P_e^H$.

We discretize the time evolution of $\rho_0$ and $\rho_1$ into equal time grids based on \eref{eq:rho_evolution}. Here, we define $u_{k,n}$ as the control amplitude in the $k$th control channel and the $n$th time slice, represented as $u_{k,n}\equiv u_k(n\Delta t)$.

The GRAPE method begins by initializing a guess for the control field $\{u_{k,n}\}$ before the GRAPE iteration. In the first iteration step, we calculate the gradient of $\D$ with respect to the control field, ${\delta\D}/{\delta u_{k,n}}$ (an analytical expression detailed in \ref{sec:grape_gradient}). Next, we update the control field by multiplying the gradient with a learning rate $\epsilon$ determined by steepest gradient descent \cite{curry1944method}. These steps are repeated until the difference between two consecutive values of $\D$ reaches a desired threshold. The pseudocode for the GRAPE algorithm is summarized in Algorithm~\ref{alg:grape}.

\begin{algorithm}[tbh]
    \SetNlSkip{0.4em}
    \SetInd{0.5em}{1em}
    \SetAlgoVlined
    \SetVlineSkip{0.5em}
    \DontPrintSemicolon
        Set initial values of the control field $\{u_{k,n}\}$\;
        \While{$\D$ $\mathrm{is}$ $\mathrm{not}$ $\mathrm{converged}$}
        {
            calculate the time evolution of $\rho$\;
            calculate $\D$ at the target time $T$\;
            calculate the gradient ${\delta\D}/{\delta u_{k,n}}$\; 
            $u_{k,n} \leftarrow u_{k,n} + \epsilon\,{\delta\D}/{\delta u_{k,n}}$\;
            \tcp{$\epsilon$ is learning rate determined by steepest gradient descent}
        }
    \caption{GRAPE}
    \label{alg:grape}
\end{algorithm}
\begin{algorithm}[tbh]
    \SetNlSkip{0.4em}
    \SetInd{0.5em}{1em}
    \SetAlgoVlined
    \SetVlineSkip{0.5em}
    \DontPrintSemicolon
    Set initial values of the control field $\{u_{k,n}^{0}\}$\;
    Set the temperature $T^0$ and the cooling factor $\alpha$\;
    \While{$\D$ $\mathrm{is}$ $\mathrm{not}$ $\mathrm{converged}$}
    {
        \For{$i\leftarrow 0$ \KwTo $\kappa$}{
            randomly choose a new control field $\{u'_{k,n}\}$\;
            calculate $\delta\D=\D'-\D^i$ at the target time\;
            \eIf(\tcp*[h]{$\Delta^i=-\min[1,T^i\exp(\delta\D/T^i)]$}){$\delta\D\geqslant\Delta^i$}{
                $\{u_{k,n}^{i+1}\} \leftarrow \{u'_{k,n}\}$
            }{
                $\{u_{k,n}^{i+1}\} \leftarrow \{u_{k,n}^{i}\}$
            }
            $T^{i+1} \leftarrow \alpha\,T^{i}$
        }
        run GRAPE iteration using $\{u_{k,n}^{\kappa}\}$\;
        $\{u_{k,n}^{0}\} \leftarrow \{u_{k,n}^{\kappa}\}$
    }
    \caption{SAGRAPE}
    \label{alg:sagrape}
\end{algorithm}

The SAGRAPE method combines the simulated annealing process with  GRAPE. Before each GRAPE iteration, the control field $\{u_{k,n}\}$ is initialized using the simulated annealing technique. This hybrid approach allows for escaping local minima and speeding up the local search near a stationary point \cite{yiu2004hybrid}.

In the simulated annealing process, an annealing schedule is defined, inspired by the heating and cooling of metals \cite{kirkpatrick1983optimization}. The schedule includes parameters such as the cooling speed $\alpha$, the number of cooling steps $\kappa$, and the initial temperature $T^0$. 
At each step of the process, random perturbations on the control field $\{u_{k,n}\}$ are generated, resulting in a modified field $\{u'_{k,n}\}$. 
The difference in the objective function, $\delta\D=\D'-\D$, is then calculated. 

The decision to accept or reject the modified control field $\{u'_{k,n}\}$ is based on the Boltzmann probability distribution \cite{kirkpatrick1983optimization,vcerny1985thermodynamical}.
If $\delta\D$ satisfies $\mathrm{Random}(0,1)<T^i\exp(\delta\D/T^i)$, where $T^i$ represents the temperature analogously to metals, the control field for the next iteration is updated to $\{u'_{k,n}\}$. Otherwise, the control field remains the same.

Additionally, in the threshold-based update,  $\{u'_{k,n}\}$ is accepted if $\delta\D\geqslant\Delta^i$, where $\Delta^i=-\min[1,T^i\exp(\delta\D/T^i)]$ serves as the threshold function. Importantly, in both cases, a random perturbation can be accepted even when  $\delta\D<0$, enabling the simulated annealing to avoid local optima.

To summarize, Algorithm~\ref{alg:sagrape} provides the pseudocode of the SAGRAPE algorithm, adapted from Ref.~\cite{ram2022robust}.

To compute the gradient for $N$ piecewise constant controls, the time evolution of the density matrix needs to be evaluated $N^2$ times in each step of the GRAPE iteration. However, it is possible to reduce the time complexity of GRAPE from $\mathcal{O}({N^2})$ to $\mathcal{O}({N})$ by allocating more computer memory to store the superoperators. In this work, the superoperators are numerically calculated using the default algorithms provided by the QuTiP package \cite{Johansson2012,Johansson2013}.

In the SAGRAPE simulations, unless explicitly stated otherwise, we use  $\alpha=0.9$, $\kappa=50$ and $T^0=0.02$ as the parameters for the simulated annealing process. 
The convergence efficiency of SAGRAPE compared to GRAPE is elaborated in Sec. \ref{sec:converge}.
The results demonstrate that, in most cases, SAGRAPE exhibits superior convergence efficiency compared to GRAPE. However, it should be noted that the SAGRAPE method does not further reduce the error probability.

It is worth mentioning that there are other numerical methods available to derive optimal controls. For example, the Krotov's method\cite{goerz2019krotov,reich2012monotonically,palao2003optimal} and the gradient-free chopped random-basis (CRAB) method \cite{machnes2018tunable,caneva2011chopped} are alternative approaches for finding optimal control strategies without relying on gradient information.

\section{Results}\label{sec:results_}

\subsection{Environmental noise}\label{sec:results_environment}

Without loss of generality, we consider the unknown signal of the magnetic field aligned along the $z$ direction.
The binary hypotheses in this case are clearly defined as $\{0,\sigma_z\}$, where $H_0$ corresponds to the hypothesis of a zero magnetic field ($0$), and $H_1$ corresponds to the hypothesis of a magnetic field $B$ along the $z$ direction ($\sigma_z$).
The control Hamiltonian is expressed in the $\sigma_x$ and $\sigma_y$ channels as follows:
\begin{equation}\label{eq:Hc-sxsy}
    H_{c}(t)=u_{x}(t) \sigma_{x}+u_{y}(t) \sigma_{y},
\end{equation}
where $\sigma_x,\sigma_y\in{SU}(2)$ ensures the controllability of the qubit system \cite{domenico2021}.
We set the initial state as an eigenstate of $\sigma_x$, $\ket{+} = \frac{1}{\sqrt{2}}(\ket{0} + \ket{1})$. 
Under the dynamics of $H_1$, the state undergoes a rotation within the $xy$ (equatorial) plane of the Bloch sphere. As a result, at the target time $T=n\pi/2$, the final state becomes orthogonal to $\ket{+}$, resulting in zero-error hypothesis testing. In this scenario, the fixed local measurement is chosen with the projective measurement elements $E_{0}=\dyad{+}$ and $E_{1}=\dyad{-}$.

\textit{Example 1: Dephasing dynamics.}
The first example studies the environmental noise of dephasing dynamics which represents a loss of coherence in a quantum system.
In this case, the two quantum channels under control are described by the following differential equation:
\begin{equation}
    \partial_t\rho_j = -i\left[H_j + H_{c}, \rho_{j}\right] + \frac{\gamma}{2}\left(\sigma_{\hat{n}}\rho_{j}\sigma_{\hat{n}} - \rho_{j}\right).
\end{equation}
Here, $j=0,1$ represents the two hypotheses. The term $\sqrt{\frac{\gamma}{2}}\sigma_{\hat{n}} \equiv \sqrt{\frac{\gamma}{2}}\vec{\sigma}\cdot\hat{n}$ is the collapse operator that induces dephasing dynamics, with $\gamma$ being the dephasing rate.
The direction of dephasing is determined by the unit vector $\hat{n} = (\sin\theta \cos\phi, \sin\theta \sin\phi, \cos\theta)$, where $\theta\in [0,\pi)$ and $\phi\in [0,2\pi)$. 
In the Bloch sphere representation, the evolved Bloch vector along $\hat{n}$ direction remains constant while other components of the Bloch vector orthogonal to $\hat{n}$ decay exponentially in time, reflecting the tendency to approach a mixed state \cite{liu2017control}.
In the subsequent analysis, we assume $\phi=0$ without loss of generality and focus on two specific cases: parallel dephasing ($\theta=0$) and transverse dephasing ($\theta=\pi/2$).

\begin{figure}[t]
    \centering
    \captionsetup{width=\linewidth}
    \includegraphics[width=0.98\linewidth]{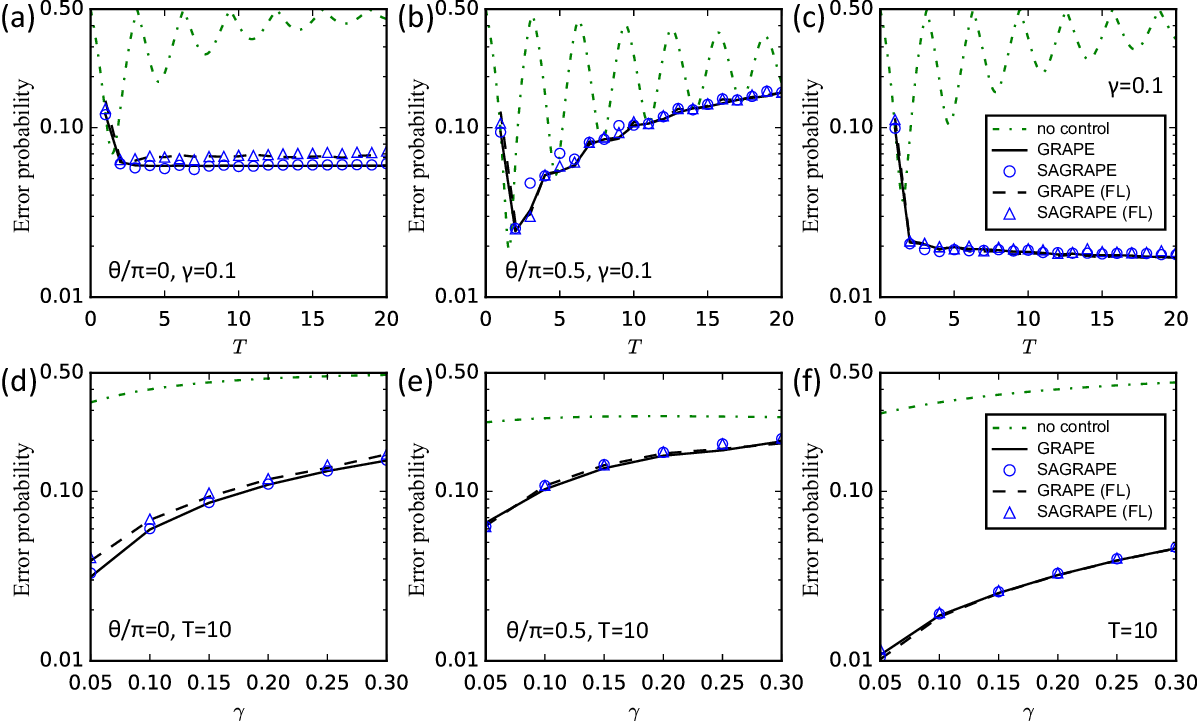}
    \caption{The error probability under dephasing dynamics and spontaneous emission.
  Panels (a) and (d) correspond to parallel dephasing, panels (b) and (e) represent transverse dephasing, while panels (c) and (f) illustrate spontaneous emission. In panels (a)-(c), the error probability is shown as a function of the target time $T$ at a dephasing rate of $\gamma=0.1$. In panels (d)-(f), the error probability is depicted as a function of the dephasing rate $\gamma$ at a fixed target time of $T=10$. The solid lines and blue circles represent the error probability $P_e^H$ under controls found using the GRAPE and SAGRAPE methods, respectively, when a Helstrom measurement is employed. The green dash-dot lines indicate $P_e^H$ under free evolution (``no control''). The black dashed lines and blue triangles depict the error probability $P_e$ obtained when a fixed local measurement is employed (remarked in the legend as ``FL'').
  }\label{fig:PvsT}
\end{figure}

We begin our discussion by examining hypothesis testing under parallel dephasing conditions. The optimal controls are obtained for various values of the parameter $T$, and the corresponding error probabilities $P_e^H$ are plotted as a function of $T$ in \fref{fig:PvsT}(a).

In the absence of control (uncontrolled scheme, shown as ``no control''), the green dash-dotted line represents $P_e^H$, which reaches its highest value of $0.5$ at $T=n\pi$ and local minimum at $T=n\pi/2$. However, for sufficiently large values of $T$, both $\rho_0$ and $\rho_1$ evolve towards the completely mixed state $I/2$, causing the minimum value of $P_e^H$ to increase to $0.5$.

On the other hand, under optimal control, $P_e^H$ decreases as $T$ increases, and it stabilizes at around $0.06$, which is approximately $1/8$ of the value for the uncontrolled case. This observation indicates that during the early stages of time evolution, $\rho_j$ reaches steady states on the $z$ axis of the Bloch sphere, where the states remain unchanged under both dephasing and external fields \cite{basilewitsch2020optimally}.

Furthermore, \fref{fig:PvsT}(d) depicts the optimal error probability as a function of the dephasing rate $\gamma$ at a fixed $T=10$. As expected, increasing $\gamma$ leads to an increase in $P_e^H$. 
For instance, the optimized value of $P_e^H$ at $\gamma=0.3$ is 0.152, approximately five times higher than $P_e^H=0.031$ at $\gamma=0.05$.
It is worth noting that the results obtained from the GRAPE and SAGRAPE methods exhibit comparable performance, as depicted by the black lines and blue markers in \fref{fig:PvsT}(a) and \fref{fig:PvsT}(d).

\begin{figure}[tbh]
    \centering
    \captionsetup{width=\linewidth}
    \includegraphics[width=0.9\linewidth]{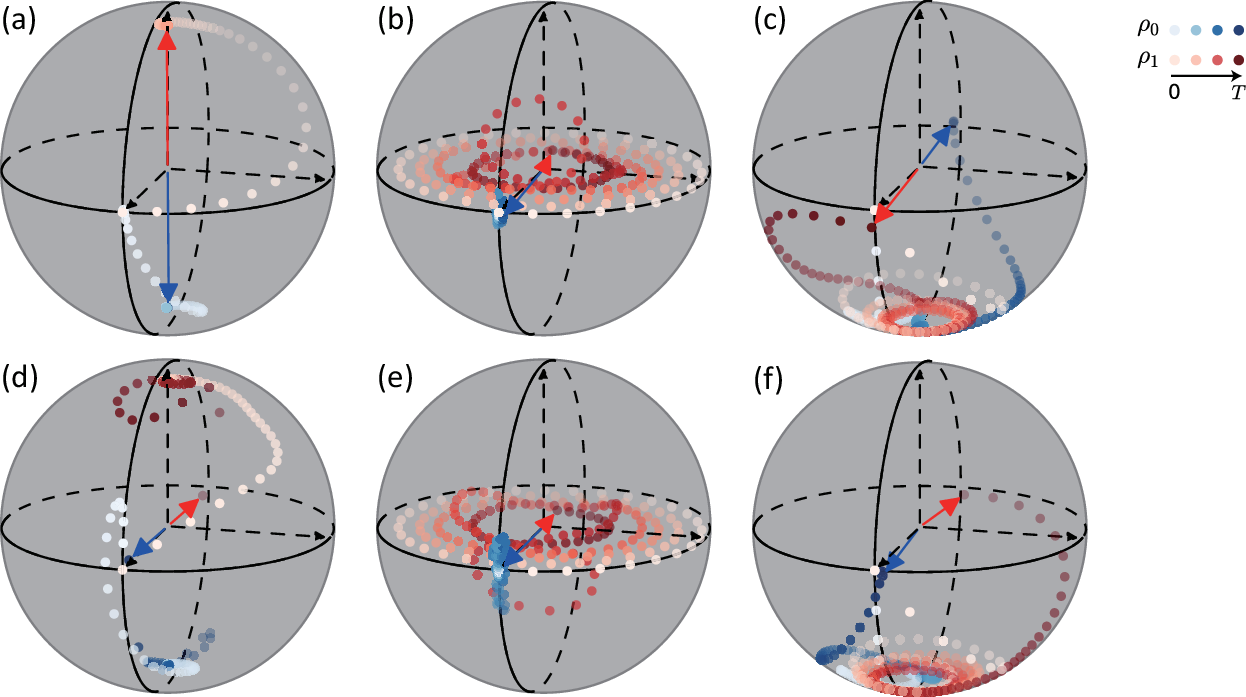}
    \caption{The trajectories of quantum states on the Bloch sphere are shown for the null and alternative hypotheses with optimal controls. The Helstrom measurement is used in (a)-(c), while the fixed local measurement is used in (d)-(f). The cases of parallel dephasing (Panels (a,d)), transverse dephasing (Panels (b,e)), and spontaneous emission (Panels (c,f)) are shown in different columns. The dephasing rate is set to $\gamma=0.1$ and the target time is $T=19$. Data points with darker colors correspond to the quantum states at later time $t$. The blue and red arrows denote the final states $\rho_0(T)$ and $\rho_1(T)$, respectively.
    }\label{fig:trajectory}
\end{figure}

In this simple example, there is an evident gap in the optimal error probabilities between the Helstrom measurement and the fixed local measurement.
In \fref{fig:trajectory}(a) and \fref{fig:trajectory}(d), we plot the trajectories of quantum states on the Bloch sphere for the Helstrom and fixed local measurements, which provides an intuitive picture for the origin of this gap.
The type of measurements leads to the distinct objective functions defined with the Helstrom measurement ($P_e^H$) and fixed local measurement ($P_e$) for the control field optimization. These different objectives may bring about varying optimal control fields and consequently different trajectories on the Bloch sphere.
As discussed in Ref.~\cite{basilewitsch2020optimally}, since the steady states under the parallel dephasing are located on the $z$ axis,
the optimal control strategy separates the two states $\rho_0$ and $\rho_1$ towards the opposite directions on the $z$ axis. 
Hence, the Helstrom measurement at $T$ is given by $\{\dyad{0},\dyad{1}\}$. 
However, for the fixed local measurement, extra rotations towards the $\ket{+}$- and $\ket{-}$-directions of the $x$ axis must be performed before measurement, 
which leads to further dephasing and higher optimal values, i.e., $P_e>P_e^H$.

We now turn to the situation of transverse dephasing. 
Since the Lindblad operator $\sigma_x$ keeps the $x$ projection of the quantum states invariant, 
$\rho_0$ of the null hypothesis is time-independent and noise-free. 
    However, the alternative hypothesis's $\rho_1$ undergoes rotation away from the $x$ axis on the equatorial plane due to the external field $\sigma_z$. Consequently, under the influence of dephasing, $\rho_1$ gradually evolves towards the state $I/2$ (i.e., the center of the Bloch sphere). This evolution exerts a crucial impact on the distinguishability of the two states. Specifically, at $T=n\pi/2$ the states $\rho_0$ and $\rho_1$ are no longer orthogonal, and at $T=n\pi$ the two states become distinguishable, contrast to their initial indistinguishability.
In \fref{fig:PvsT}(b), $P_e^H$ of the uncontrolled scheme consequently oscillates as a function of $T$, 
and the local maxima (minima) decays (grows) towards $P_e^H=0.25$ when $T$ is increased.
Although the optimal control minimizes $P_e^H$ and suppresses the oscillation of $P_e^H$ in general, the error probability at $T=n\pi/2$ is comparable with that of the uncontrolled scheme.
This means that the transverse dephasing and the external field are intertwined with each other,
and there is no clear trade-off between them for optimal control. 
In addition, when $\gamma$ increases from $0.05$ to $0.3$, the optimal error probability at $T=10$ is increased from 0.065 to 0.197 (about three times higher), as shown in \fref{fig:PvsT}(e).

Interestingly, the optimal error probabilities are comparable between the fixed local measurement and the Helstrom measurement, which is very different from the parallel dephasing case.
As in the parallel dephasing case, representative trajectories of time-evolved states are presented in \fref{fig:trajectory}(b) and \fref{fig:trajectory}(e) for the Helstrom and fixed local measurements, respectively.
The time-evolved $\rho_0$ is restrained at the initial state $\dyad{+}$, but $\rho_1$ evolves on the equatorial plane under controls. 
Intuitively, the key point of the optimal control is to generate a $\sigma_x$-rotation that lead to a phase shift $\pi$ about the $z$ axis.
After some evolution, $\rho_1(T)$ is then relocated along the $\ket{-}$-direction of the $x$ axis.
At this point, the fixed local and Helstrom measurements are equivalent, and thus the optimal error probabilities satisfy $P_e\approx P_e^H$.


\textit{Example 2: Spontaneous emission.}
The second example analyzes the spontaneous emission, which is another common source of environmental noise in quantum systems.
In this case, the master equation reads
\begin{eqnarray}
        &\partial_t\rho_j = &-i\left[H_j + H_{c}, \rho_{j}\right] + \frac{\gamma_{-}}{2}\left(2\sigma_{-}\rho_{j}\sigma_{+} - \{\sigma_{+}\sigma_{-},\rho_{j}\}\right) \nonumber \\
        &&+ \frac{\gamma_{+}}{2}\left(2\sigma_{+}\rho_{j}\sigma_{-} - \{\sigma_{-}\sigma_{+},\rho_{j}\}\right),
\end{eqnarray}
where $\sigma_{\pm}=(\sigma_x \pm i\sigma_y)/2$. For simplicity, we set $\gamma_{-}=\gamma$ and $\gamma_{+}=0$. 
The Lindblad operator $\sigma_{-}$ quickly drives the initial state $\dyad{+}$ away from the equatorial plane.
In particular, $\rho_0$ of the null hypothesis evolves towards the ground state $\dyad{0}$ in the $xz$ plane, while for the alternative hypothesis, $\rho_1$ is rotated by the external field $\sigma_z$ about the $z$ axis besides evolving towards $\dyad{0}$.
Hence, $P_e^H$ without control oscillates while increasing $T$, as shown in \fref{fig:PvsT}(c).
In comparison, $P_e^H$ of the controlled scheme quickly  stabilizes at 0.017, a value that is $1/20$ of that of the uncontrolled case ($P_e^H=0.332$ at $T=20$).
Moreover, \fref{fig:PvsT}(f) shows the optimal error probability versus the decay rate $\gamma$ at $T=10$.
As in the dephasing cases, when $\gamma$ increases from $0.05$ to $0.3$, the optimal error probability is increased from 0.011 to 0.046, i.e. by four times.

The Helstrom measurement, once again, does not have the advantage over the fixed local measurement in respect of the optimal error probability.
In \fref{fig:trajectory}(c) and \fref{fig:trajectory}(f), we plot the trajectories of $\rho_0,\rho_1$ on the Bloch sphere for the Helstrom and fixed local measurements, respectively.
A similar situation has been discussed in Ref.\cite{basilewitsch2020optimally}. In their model, the free Hamiltonians only differ by a tiny magnetic field and thus $\rho_0,\rho_1$ can be simultaneously transformed (almost) to the steady state $\dyad{0}$ under control.
But in our case, even if the control transforms $\rho_0$ to $\dyad{0}$, 
the alternative hypothesis $\sigma_z$ makes it impossible to map $\rho_1$ to $\dyad{0}$.
In fact, while $\rho_0$ is transformed to $\dyad{0}$, $\rho_1$ evolves on a small circle near the $\ket{0}$-pole, as shown in \fref{fig:trajectory}(c) and \fref{fig:trajectory}(f).
Finally, $\rho_0,\rho_1$ are driven away from $\dyad{0}$ and are changed back onto the equatorial plane, pointing to any opposite directions at the target time, such that $\D_{\tr}$ becomes maximal \cite{basilewitsch2020optimally}.
The fixed local measurement are simply constructed within all opposite directions on the equatorial plane at the target time, and thus we have $P_e\approx P_e^H$.

\begin{figure}[t]
    \centering
    \captionsetup{width=\linewidth}
    \includegraphics[width=0.98\linewidth]{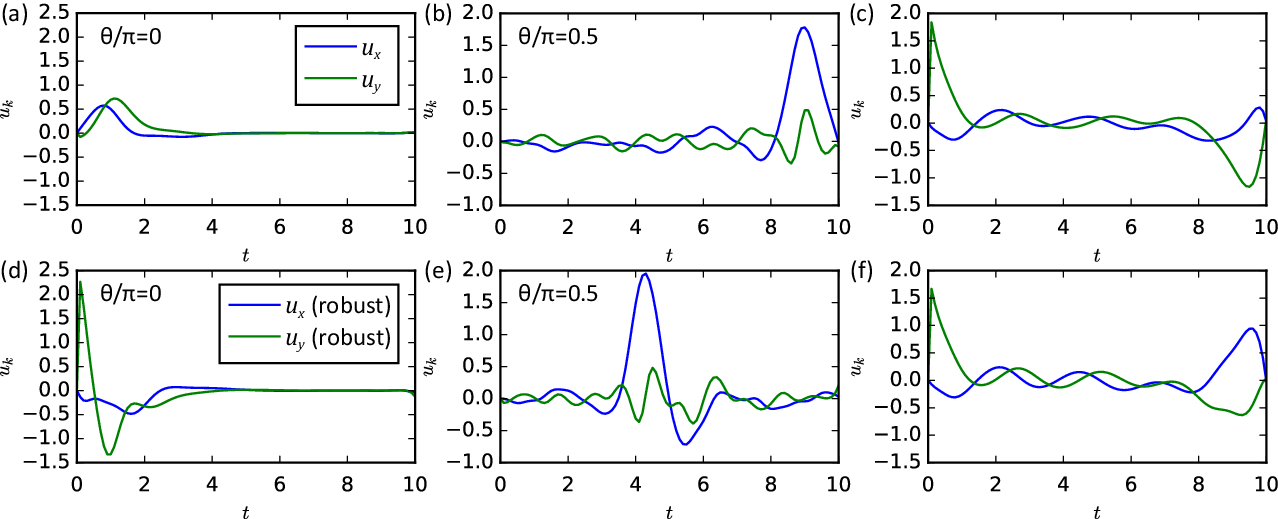}
    \caption{The pulse profiles optimized with or without the signal noise using the GRAPE method. In (a)-(c), the pulse profiles are for the perfect signal, while in (d)-(f), the pulse profiles are designed to have robustness against the signal noise $d\omega\in[-0.1,0.1]$. 
    The cases of parallel dephasing (Panels (a,d)), transverse dephasing (Panels (b,e)), and spontaneous emission (Panels (c,f)) are shown in different columns.
    The target time for the optimization is set to $T=10$.
    }\label{fig:robust_control}
\end{figure}

\subsection{Imperfect signal}\label{sec:results_signal}
In the previous section, we have numerically employed the optimal control in quantum hypothesis testing 
given the nominal value of an unknown external field.
However, in a real-world physical system, the external field can be compromised by the imperfect signal, deviating from the nominal value.
In this section, we address the problem of the robust schemes that are resilient to the signal noise inside a tolerance window.

Consider the signal noise in the dimension of frequency of the external field.
In hypothesis testing, the free Hamiltonians can be written as $H_0 = 0$ and $H_1=(1+d\omega)\sigma_z$ where $d\omega$ is drawn from a uniform distribution.
We focus on the average error probability over an interval $d\omega\in[-L/2,L/2]$, 
and thus the objective function is defined as
\begin{equation}\label{eq:perrave}
    \expval{P_e^H} = \frac{1}{L}\int_{-L/2}^{L/2} P_e^H\,d\omega,
\end{equation}
where $L$ represents the length of the tolerance window $[-L/2,L/2]$, with larger $L$ indicating a greater deviation of system Hamiltonian parameters from their nominal values.
In our simulations, we uniformly choose $21$ samples within the interval $d\omega\in[-0.1,0.1]$ to evaluate the gradient using \eref{eq:perrave} and optimize the control field accordingly.
Since the GRAPE and SAGRAPE methods give comparable results in the previous section, we therefore only discuss the robust control using the GRAPE method.

First, representative pulse profiles of the optimal control for perfect signal and the robust control are depicted in \fref{fig:robust_control}(a)-(c) and \fref{fig:robust_control}(d)-(f), respectively.
In the parallel dephasing case, the robust control has a stronger $u_y$ kick in the early stage of time evolution, compared to the optimal control, as shown in \fref{fig:robust_control}(a) and \fref{fig:robust_control}(d).
For transverse dephasing, the optimal control tunes the course of evolution near the target time $T$, designated by the higher crest in $u_x,u_y$ in \fref{fig:robust_control}(b).
By contrast, in \fref{fig:robust_control}(e) the peak of robust control acts at around $T/2$. 
For spontaneous emission, as can be seen in \fref{fig:robust_control}(c), the $u_x,u_y$ pulse peak in the beginning and the end of time evolution where the first peak transforms $\rho_0,\rho_1$ close to $\dyad{0}$ and the second one changes the states back to the equatorial plane, similar to those in Ref.~\cite{basilewitsch2020optimally}.
The robust control, on the other hand, has almost the same pulse shape except for
having stronger $u_x$ and weaker $u_y$ components near $T$, as shown in \fref{fig:robust_control}(f).

\begin{figure}
    \centering
    \captionsetup{width=\linewidth}
    \includegraphics[width=0.98\linewidth]{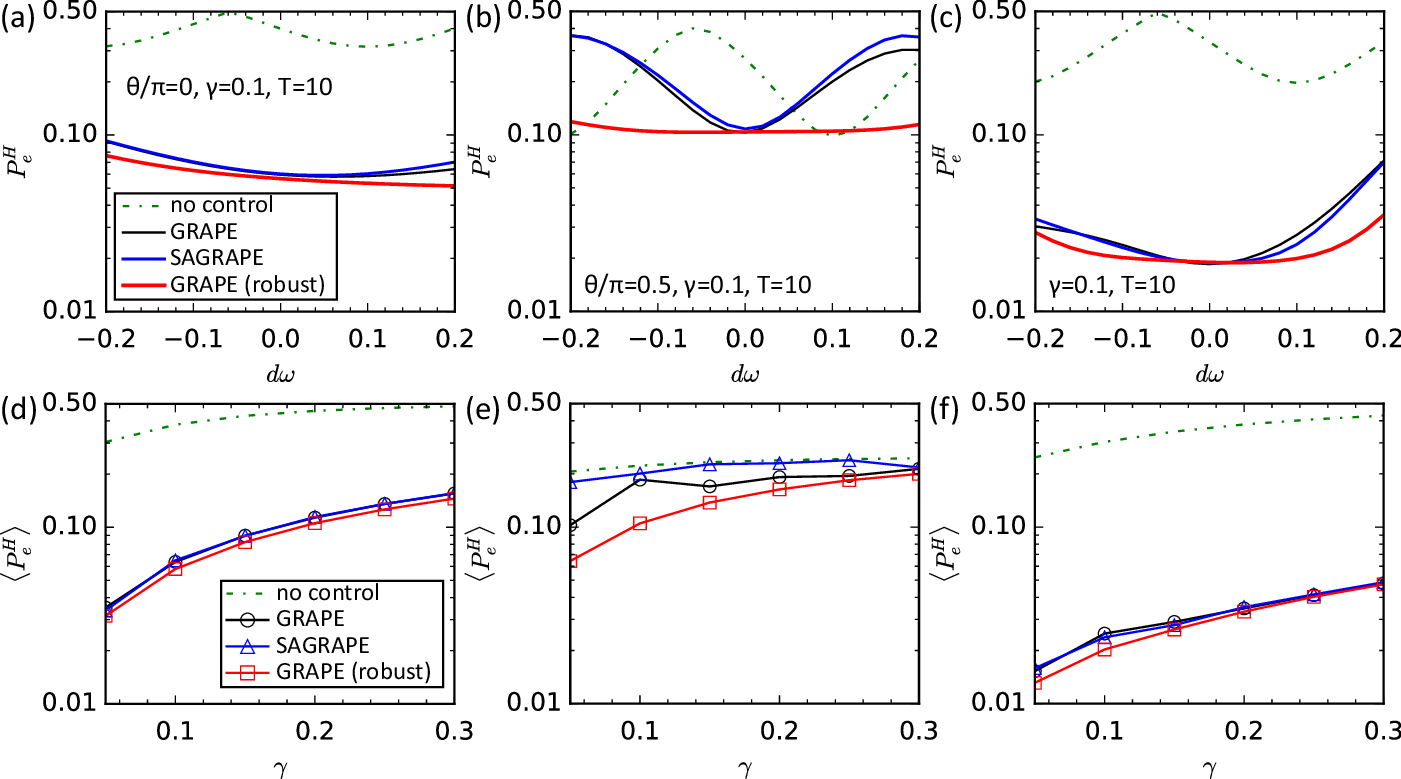}
    \caption{Robustness of the optimal controls in the presence of signal noise.
 In (a)-(c), the error probability, $P_e^H$, is shown as a function of the signal noise $d\omega$, which is evenly sampled in the range $[-0.1,0.1]$. In (d)-(f), the average error probability, $\expval{P_e^H}$, over the interval $[-\pi/2T,\pi/2T]$ is plotted as a function of the dephasing rate $\gamma$. The target time for the optimization is set to $T=10$.
        }\label{fig:robust}
\end{figure}
    
Next, we test the optimal control and the robust control when the signal noise $d\omega\in[-0.2,0.2]$. 
The robustness of control, $\expval{P_e^H}$, is systematically evaluated for different $\gamma$ by the integration over $d\omega\in[-\pi/(2T),\pi/(2T)]$ in \eref{eq:perrave}.

In the case of parallel dephasing, the error probabilities as a function of $d\omega$ at $T=10,\gamma=0.1$ are presented in \fref{fig:robust}(a).
The optimal control and the robust control perform equally well at $d\omega=0$, and the robust control slightly reduces the error probability when $d\omega$ is large. 
We have shown that under optimal control $\rho_0,\rho_1$ quickly evolve to the steady states to minimize the effect of dephasing. In this case, $\rho_0,\rho_1$ on the $z$ axis are stable against the external field.
Thus, the optimal control at $d\omega=0$ intrinsically has a level of robustness.
On the other hand, for transverse dephasing in \fref{fig:robust}(b), $P_e^H$ under optimal control is a periodic function of $d\omega$, leading to an error probability worse than the uncontrolled scheme, 
while a vast improvement in distinguishability extends beyond the range of $[-0.1,0.1]$, i.e. the predefined tolerance window for control optimization, to a larger interval under robust control.
Based on the pulse profiles in \fref{fig:robust_control}(e), we argue that the robust control makes differences at around $T/2$, 
so that the second half of the evolution can be utilized to suppress the effect of signal noise, and the desired unitary transformations are accumulated for $\rho_1$.
However, under optimal control the $\sigma_x$-rotations are fixed at the end of evolution, so the signal noise can significantly influence $\rho_1(T)$ and thus $P_e^H$.

The similar analysis for spontaneous emission is carried out in parallel. 
In \fref{fig:robust}(c), we plot $P_e^H$ versus $d\omega$ at $T=10,\gamma=0.1$.
A moderate improvement is seen when using the robust control.
More precisely, $P_e^H$ is almost a constant for the noise $d\omega\in[-0.1,0.1]$, while $P_e^H$ under optimal control increases much faster when $|{d\omega}| > 0$.
We recall that at the end of time evolution, $\rho_0,\rho_1$ around the $\ket{0}$-pole of the Bloch sphere have to be transformed onto the equatorial plane for measuring.
For the perfect signal $d\omega=0$, the final states $\rho_0(T),\rho_1(T)$ are along the opposite directions,
while with the signal noise $d\omega>0$, the trajectories of time-evolved $\rho_1$ cannot lead to the optimal $\rho_1(T)$, giving rise to higher $P_e^H$.
On the other hand, the robust control, by adjusting $u_x,u_y$ components near the target time, may identify the trajectories that minimize the effect of signal noise during the rotation towards the equatorial plane.

We also calculate $\expval{P_e^H}$ in a larger interval $d\omega\in[-\pi/(2T),\pi/(2T)]$ as a function of $\gamma$ for different noisy dynamics.
As can be seen in \fref{fig:robust}(d)-(f), 
the controlled schemes have much smaller $\expval{P_e^H}$ than the case without control. 
Among them, the robust control gives the lowest $\expval{P_e^H}$.
Compared to the optimal control for perfect signal, the robust control further reduces $\expval{P_e^H}$ and shows the better robustness.
In particular, for transverse dephasing with $\gamma=0.1$, the robust control further cuts $\expval{P_e^H}$ down about $40\%$.
Nonetheless, the improvement brought by the robust control decreases with increasing the dephasing rate or decay rate $\gamma$, as shown in \fref{fig:robust}(e).

\section{Conclusions and outlook}\label{sec:conclusion}
We have utilized optimal control methods, specifically the GRAPE and SAGRAPE algorithms, for quantum hypothesis testing in the presence of environmental noise. Our study includes a comparison between scenarios with and without considering the impact of experimental imperfections in the signal frequency.
For parallel dephasing and spontaneous emission, the optimal control inherently exhibits a certain level of robustness. However, in the case of transverse dephasing with an imperfect signal, the optimal control may result in a higher error probability compared to the uncontrolled scheme.
In contrast, the robust control, optimized considering a range of signal noise, demonstrates superior robustness even beyond the predefined tolerance window.
On average, both the optimal control and robust control exhibit improvements over the uncontrolled schemes for various dephasing or decay rates, with the robust control yielding the lowest error probability.

The findings presented in this study contribute to the understanding of robust quantum  control \cite{koswara2021quantum}
through discriminating dynamics.
There are potential future developments in the area of control robustness that could include exploring other sources of noise, such as different types of signal noises or disturbances in the pulse profiles \cite{timoney2008error,grace2012optimized,goerz2014robustness, koswara2014robustness,koswara2021robust}.
These challenges arise due to the practical difficulties in accurately implementing theoretical control designs in the laboratory \cite{brif2010control}. 
Another approach to achieving robustness is through the application of the Pontryagin maximum principle \cite{domenico2021}. These investigations hold promise for the development of robust quantum hypothesis testing, which not only is valuable for theoretical research but also has practical implications for laboratory control.

The multi-objective control problem \cite{brif2010control} has been addressed using the GRAPE and SAGRAPE methods, aiming to simultaneously reduce both type I and type II errors.  These techniques are also crucial for optimizing robust control, where a weighted-sum objective functional \eref{eq:perrave} is typically employed.
In recent years, multi-objective tasks have been increasingly approached using a machine learning technique called the
multi-task learning \cite{caruana1997multitask,ruder2017overview}.
This raises the question of whether machine learning approaches can achieve robust control pulses comparable to GRAPE for quantum hypothesis testing.
Furthermore, instead of focusing on binary hypothesis testing in single-qubit systems, one can explore optimal control strategies for multiple-qubit states. This would involve quantum control of entangled qubits \cite{piani2009all}, which may lead to lower optimal error probabilities determined by the quantum Chernoff bound \cite{audenaert2007discriminating}.

\section*{Acknowledgments}
H.X. and B.W. contribute equally to this work. This work is supported by the Key-Area Research and Development Program of GuangDong Province  (Grant No. 2018B030326001), the National Natural Science Foundation of China (Grant No. 11874312), the Research Grants Council of Hong Kong (Grant No. CityU 11303617, CityU 11304018, CityU 11304920), and the Guangdong Innovative and Entrepreneurial Research Team Program (Grant No. 2016ZT06D348). H.Y. acknowledges partial support from the Research Grants Council of Hong Kong with Grant No. 14307420, 14308019, 14309022.

\section*{Code availability}
The source code is available upon reasonable request from the authors.

\section*{Data availability}
The data that support the findings of this study are available upon reasonable request from the authors.

\appendix
\section{Gradient calculation in GRAPE}\label{sec:grape_gradient}
According to \eref{eq:obj_func_fix} and \eref{eq:obj_func}, the object functions are
\begin{equation}
    -P_e = - \pi_0\tr(\rho_0E_1) - \pi_1\tr(\rho_1E_0),
\end{equation}
where $E_0$ and $E_1$ are the fixed POVM elements, and
\begin{equation}
    \D(\rho_0, \rho_1) = \frac{1}{2}\tr[\left(\rho_0-\rho_1\right)^{\dagger}\left(\rho_0-\rho_1\right)].
    \label{eq:obj_func_appendix}
\end{equation}
Calculating the gradients with respect to the control amplitude $u_k$, we obtain,
\begin{equation}
    -\pdv{P_e}{u_{k,n}}
    = -\pi_0\tr (\pdv{\rho_{0}}{u_{k,n}}E_1)-\pi_1\tr (\pdv{\rho_{1}}{u_{k,n}}E_0),
\end{equation}
and
\begin{equation}
    \pdv{\D}{u_{k,n}}
    =\frac{1}{2}\tr[(\pdv{\rho_{0}^{\dagger}}{u_{k,n}}-\pdv{\rho_{1}^{\dagger}}{u_{k,n}})(\rho_0-\rho_1)
    +\mathrm{H.c.}].
    \label{eq:D_Vj}    
\end{equation}
Here the key quantity is the derivative ${\partial \rho}/{\partial u_{k,n}}$.
Next, we define the propagation operator,
\begin{equation}
    D_{n+1}^{m}=\prod_{i=n+1}^{m} e^{\Delta t \L_i},
\end{equation}
that maps the density matrix at the time $n\Delta t$ to the time $m\Delta t$.
For example, the density matrix at the target time, $\rho^{N}\equiv\rho(N\Delta t)$, satisfies
\begin{equation}
    \rho^{N}=D_{n+1}^{N} \, \rho^{n}.
\end{equation}
Substitute the above definition to ${\partial \rho^{m}}/{\partial u_{k,n}}$, we have
\begin{equation}\label{eq:supp:dv-0}
    \pdv{\rho^{m}}{u_{k,n}}
    =\pdv{D_{n+1}^{m}}{u_{k,n}}\,\rho^{n} + D_{n+1}^{m}\,\pdv{\rho^{n}}{u_{k,n}}
    =D_{n+1}^{m}\,\pdv{\rho^{n}}{u_{k,n}},
\end{equation}
where ${\partial D_{n+1}^{m}}/{\partial u_{k,n}}=0$
since the propagator that explicitly has $u_{k,n}$ is $e^{\Delta t \L_n}$ in the $n$th time slice.
For the last term, we use $\rho^{n} = e^{\Delta t \L_n}\rho^{n-1}$, so that
\begin{equation}\label{eq:supp:dv-1}
    \pdv{\rho^{n}}{u_{k,n}}=\pdv{e^{\Delta t \L_{n}}}{u_{k,n}} \rho^{n-1}.
\end{equation}
The problem now is to find the gradient of the propagator $e^{\Delta t \L_n}$.

It is known that the derivative of an exponential operator satisfies $\partial_x e^{A(x)} = \int_0^1e^{sA}(\partial_x A)e^{(1-s)A} ds$ \cite{khaneja2005optimal}. Thus,
\begin{equation}\label{eq:supp:exponent}
    \pdv{e^{\Delta t \L_{n}}}{u_{k,n}}
    =\int_{0}^{1} e^{s \Delta t \L_{n}}\left(\Delta t \pdv{\L_{n}}{u_{k,n}}\right) e^{(1-s) \Delta t \L_{n}} ds. 
\end{equation}
Since $\L_{n}(A)=-i[H_{j}+\sum_{k} u_{k,n} H_{c,k}, A]+\Gamma(A)$,
we have ${\partial \L_{n}/ \partial u_{k,n}} = -i H_{c,k}^{\times}$,
where $H_{c,k}^{\times}$ represents the commutation superoperator, i.e., $H_{c,k}^{\times} A=[H_{c,k}, A]$.
Finally, using the Taylor expansion $e^{\pm s\Delta t \L_n}=\sum_{i}\frac{1}{i!}(\pm1)^i(s\Delta t)^i\L_n^i$, \eref{eq:supp:exponent} becomes,
\begin{eqnarray}
        &\frac{\delta e^{\Delta t \L_n}}{\delta V_{k,n}} 
        =-i \Delta t \int_{0}^{1} e^{s \Delta t \L_n} H_{c,k}^{\times} e^{(1-s)\Delta t \L_n} ds \\ 
        &=-i \sum_{i,j=0}^{\infty}  \frac{(-1)^{j}(\Delta t)^{i+j+1}}{(i+j+1)i! j!}\L_n^{i} H_{c,k}^\times \L_{n}^{j} e^{\Delta t \L_{n}}
\end{eqnarray}
where $\int_0^{1} s^{i+j} ds = {1}/{(i+j+1)}$.
We can define $K_{ij}$ as 
\begin{equation}\label{eq:K_nm}
    K_{ij} = \frac{(-1)^{j}(\Delta t)^{i+j+1}}{(i+j+1)i! j!}\L_{n}^{i} H_{c,k}^\times \L_{n}^{j},
\end{equation}
so \eref{eq:supp:dv-1} is rewritten as
\begin{equation}
    \pdv{\rho^n}{u_{k,n}}=-i \sum_{i,j} K_{ij} e^{\Delta t \L_{n}} \rho^{n-1} = -i \sum_{i,j} K_{ij} \rho^{n}.
\end{equation}
In our simulations, we keep the first three $K_{ij}$ terms up to the second order of $\Delta t$,
\begin{eqnarray}
    K_{00} &= \Delta t H_{c,k}^{\times},\\
    K_{01} &= \frac{-\Delta t^{2}}{2} H_{c,k}^{\times} \L_j,\quad
    K_{10} = \frac{-\Delta t^{2}}{2} \L_{j} H_{c,k}^{\times}.
\end{eqnarray}
Moreover, since our purpose is to maximize $-P_e$ and $\mathcal{D}$ at the target time $N\Delta t$, we should consider the derivative of the density matrix $\rho^{N}$ in \eref{eq:supp:dv-0}, i.e.,
\begin{equation}
    \pdv{\rho^{N}}{u_{k,n}}
    =-i D_{n+1}^{N}\sum_{i,j} K_{ij} \rho^{n}.
\end{equation}  
The above equation is then substituted in ${\partial P_e}/{\partial u_{k,n}}$ and ${\partial \D}/{\partial u_{k,n}}$.

\begin{figure}[t]
    \centering
    \captionsetup{width=\linewidth}
    \includegraphics[width=0.8\linewidth]{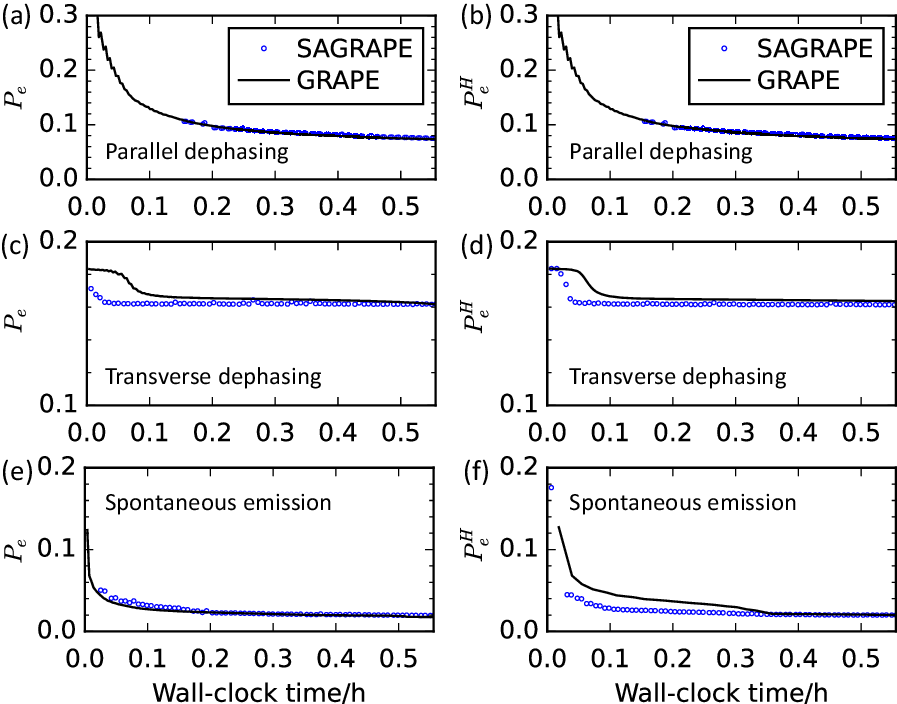}
    \caption{Convergence analysis of the GRAPE and SAGRAPE methods.
    The error probabilities $P_e$ and $P_e^H$ are shown as a function of the wall-clock time cost in hours.
        $P_e$ and $P_e^H$ under optimized controls found using the GRAPE and SAGRAPE methods are depicted by the solid lines and blue circles, respectively.
        The fixed local measurement (Panels (a,c,e)) and the Helstrom measurement (Panels (b,d,f)) are shown in different columns.
        The cases of (a)(b) parallel dephasing, (c)(d) transverse dephasing, and (e)(f) spontaneous emission are shown in different rows.
        }\label{fig:converge}
\end{figure}

\section{Convergence analysis of the GRAPE and SAGRAPE methods}\label{sec:converge}

According to a prior work \cite{ram2022robust}, 
the SAGRAPE method with varying numbers of cooling steps $\kappa$ significantly improves the convergence efficiency in both state and gate control tasks.
In this section, we demonstrate the convergence of the GRAPE and SAGRAPE methods
for the hypothesis testing problem, as discussed in the main text.

In \fref{fig:converge}, the black solid lines represent the results of GRAPE,
and the blue circles are obtained by one step of SAGRAPE iteration and one step of GRAPE iteration consecutively.
We have used the pulse profile $u_x(t)=u_y(t)=0.01$ as the initial values of control field $\{u_{k,n}\}$.
In our work, we apply the steepest gradient descent for the learning rate $\epsilon$ of the GRAPE method,
which greatly improves the convergence efficiency.
For parallel dephasing, the GRAPE and SAGRAPE methods show comparable convergence efficiency.
By constrast, the SAGRAPE method shows a better convergence efficiency in tasks under transverse dephasing and spontaneous emission. In addition, both methods converge to the same optimal error probability after long computing time.

\section*{References}

\providecommand{\newblock}{}

\end{document}